\documentclass{aa}
\usepackage{graphicx}

\begin{document}

\title{An XMM-Newton Study of the sub-structure in  M87's halo}

\author{E. Belsole\inst{1} \and J.L. Sauvageot\inst{1} \and H. B\"ohringer\inst{2} \and D.M. 
Worrall\inst{3} \and K. Matsushita\inst{2} \and R.F.  Mushotzky\inst{4} \and I. 
Sakelliou\inst{5} \and S. Molendi\inst{6} \and M. Ehle\inst{7} \and J. Kennea\inst{8} \and G. 
Stewart\inst{9}, W. T. Vestrand\inst{10}}

\authorrunning{Belsole et al.}
\offprints{E. Belsole \email{ebelsole@cea.fr}}

\institute{Service d'Astrophysique, CEA Saclay, L'Orme des Merisiers 
B\^at 709., F-91191 Gif-sur-Yvette Cedex, France. 
\and Max-Planck-Institut f\"ur Extraterrestrial Physik, D-85748 Garching, 
Germany 
\and Department of Physics, University of Bristol, Tyndall Avenue, Bristol BS8 
1TL, UK.
\and Laboratory of High Energy Astrophysics, Code 660, NASA/Goddard Space Flight 
Center, Greenbelt, MD 20771, USA
\and Mullard Space Science Laboratory, University College London, Holmbury St 
Mary, Dorking, Surrey RH5 6NT
\and Istituto di Fisica Cosmica, via Bassini 15, I-20133, Milano, Italy
\and XMM-Newton SOC, Villafranca, Apartado 50727, E-28080 Madrid, Spain
\and University of California, Santa Barbara, USA
\and Departement of Physics and Astronomy, The University of Leicester, 
Leicester LE1 7RH
\and NIS-2, MS D436, Los Alamos National Laboratory, Los Alamos, NM 87545 USA.
}
\date{}

 \abstract{
The high signal to noise and good point spread function of XMM have allowed the first detailed 
study of the interaction between the thermal and radio emitting plasma in the central regions of 
M87. We show that the X-ray emitting structure, previously seen by ROSAT, is thermal in nature
and that the east and southwest extensions in M87's
 X-ray halo have a significantly lower temperature (k$T=$ 1.5 keV) than  the surrounding ambient medium 
(k$T=$ 2.3 keV). There is little or no evidence for non-thermal emission with an upper limit on the
 contribution of a  power  law component of spectral index flatter than 3 being less than  1$\%$ of the flux in the region of the radio lobes. 
}
\maketitle


\section{Introduction}

The giant elliptical galaxy M87 is located at the center of the  X-ray diffuse 
emission of the irregular Virgo Cluster (Fabricant et al. 1980). Its nearby position ($\sim$ 17-20 Mpc; Freedman et al. 1994, Tammann \& Federspiel 
1997) and the XMM-EPIC sensitivity allow detailed measurements of the spectral structure of the X-ray emitting gas on scales down to $\sim$1 kpc. 

Recently, Harris et al. (2000) using ROSAT/HRI data  and the recent
high resolution 90 cm map of Owen et al. (2000) find general, but not 
precise coincidence between some radio and X-ray emitting regions (features 
described as the eastern `ear' and the southern `cobra'). Earlier,  B\"ohringer et al. (1995) using ROSAT/PSPC  showed that the lack of detailed spatial correlation of the radio and X-ray emission and the apparent  thermal nature of the excess emission in the X-ray asymmetric structure  exclude the inverse Compton (IC) mechanism as the origin of most of the X-ray radiation.

The observed X-ray density and temperature distribution  of the large scale emission (Mushotzky \& Szymkowiak 1988 and references therein)  suggests a radiative energy loss of the order of 10$^{43}$ ergs s$^{-1}$ which implies that the gas is cooling and it was well modeled, before the XMM data, by a classical  cooling-flow model (Stewart et al. 1984, Fabian et al. 1984). In the same region of space a pair of relativistic jets is emerging from the active nucleus of M87, ejecting  energy into this environment at an estimated rate of  10$^{44}$ ergs s$^{-1}$ in the form of relativistic plasma (e.g. Owen et al. 2000). 
How these two media interact and if this leads to a heating of the 
intra-cluster plasma is still a matter of debate.

Detailed analysis of the radio data  (Owen et al. 1999,2000) 
finds a competition between inflow of the hot cluster gas and violent outflow of energy from the inner region, but the relation between the radio jet and the 
outer halo is still not fully understood.

The  detailed morphology of the radio emission suggests 
the  ``buoyant bubbles'' interpretation, as  first proposed for radio lobes in
galaxies by Gull and Northover (1973).  
Churazov et al. (2000) have recently modeled this process with specific application to M87 to explain 
both the radio lobes and the  surrounding  X-ray 
structures. In this model, the ambient gas is captured and uplifted by 
the relativistic gas; the bubble expands and transforms itself into a torus 
rising in the potential well,  and one expects to find  thermal  gas, originating in the central regions, in the cavity of the torus. 
This dynamical behavior  implies that the high surface brightness regions are from the uplifted gas and that this effect is increased by cooling-flow conditions.

XMM-Newton observations, which provide unprecedentedly good combined
spectral and imaging capabilities, well matched to the spatial scale
sizes of interest,  are of key importance in clarifying the nature
of M87's X-ray emission. 

In this paper, we focus our study on the asymmetric X-ray enhancements detected by XMM-Newton.   
The core and jet data, together with the characteristics of the underlying 
extended gas distribution 
are presented in  B\"ohringer et al. (2001) in this volume.
High spectral resolution RGS data will be published in a later paper by  
Sakelliou et al. (2001).

The paper is organized as follows: we present the observations in section 2. 
Section 3 describes the spectro-imaging data analysis. Section 4 is dedicated to 
the scientific discussion and to the relationship between the X-ray and radio 
emission. The conclusions are in section 5.
\section{Observations and data reduction}

M87 was observed with XMM-Newton (Jansen et al. 2001) in orbit  97 during the 
performance verification phase, and all the instruments   were operating. In this paper we present only 
observations from EPIC/MOS (Turner et al. 2001) and pn (Str\"uder et al. 2001).

The  MOS and pn detectors observed  M87 in Full Frame Mode for an  
effective exposure time of about 39 and  25.9 ksec respectively.
The MOS calibrated event files were obtained using an IDL software
package developed at Saclay for calibration purposes.  While results
are very similar to those obtained with the SAS software, the
flexibility of the IDL software makes it particularly well suited to
our purpose, in particular for extracting spectra in complex spatial
regions and for computing our energy map.  In addition, this software
 has a built-in vignetting correction.
The detector background, as indicated by the 
light curve of high energy (10-12 keV) events measured with the
MOS, was unusually high at the start of the observation, and decreased
linearly over the first 7 ksec.  We
consequently ignore the first 7 ksec of MOS data. Our  pn event list were derived from 
raw data by preliminary SAS tools and further analyzed by the public SAS 
software. We use the full exposure time. 

\section{Data Analysis}

\subsection{Image analysis: morphology of the extended features}

The surface-brightness enhancements at the location of the radio lobes  are embedded in the galaxy and cluster diffuse emission 
which has a steep radially falling surface brightness profile.  
To show their morphology more clearly we thus need a good model of the diffuse gas. The extended M87 X-ray halo  does not follow a simple King profile 
and deviations are seen at a distance greater than 5 arcmin in radius from the 
center (B\"ohringer et al. 1997), so modeling this emission is not trivial.

The wavelet transform  is a powerful tool for structure decomposition  on 
different image scales. 
\noindent We added together the MOS1 and MOS2 [0.2 - 10.0] 
keV images and then applied the {\it \`a trou\/}{\footnote{see Stark, J.L., 
Murtagh, F., Bijaoui, A., in \em{Image Processing and 
Data Analysis} \rm Cambridge University Press. 1998, p.21 \rm for 
details} algorithm implemented in the MR/1 wavelets package (Stark 1999).
The sum of the lowest spatial frequency scales (corresponding to 4.4 
to 17 arcmin with our pixel size{\footnote{1 pixel = $4.1''\times 4.1''$}) represents mainly  the cluster and galaxy 
diffuse gas emission. We subtracted this contribution from the original image 
revealing the high spatial frequency features (see figure 1).

\begin{figure}
\begin{center}
\resizebox{7.cm}{!}{\includegraphics{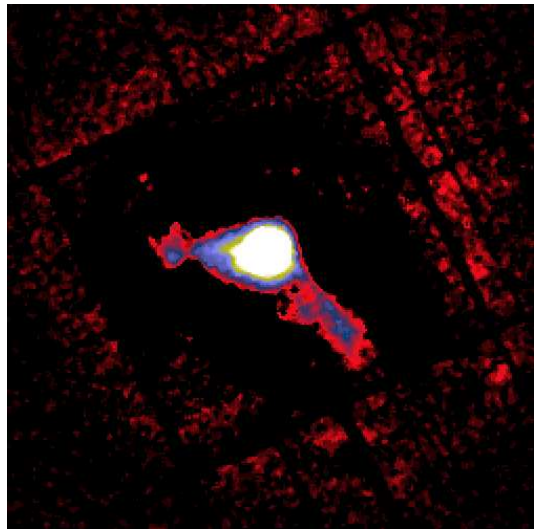}}
\caption{Combined MOS1 and MOS2 0.2-10 keV image of the asymmetric extended X-ray 
arms; north is up, east is to the left. A model of the M87 extended halo, 
obtained using the  wavelet analysis, has been subtracted.The image covers a field of view of 17 arcmin.}
\label{figure1}
\end{center}
\vspace{-0.5cm}
\end{figure}

As a first approach, we have extracted spectra in rectangular regions of nearly 3.5$\times$2 arcmin covering 
the eastern and south-western  features apparent in figure 1. Since the 
underlying halo emission is roughly symmetric, we have extracted the spectrum of the ambient medium  
in a circle of 4.5 arcmin radius, centered on the core of
 M87 and excluding the nucleus, jet  and the features associated with the 
radio lobes.
 The net spectra are clearly 
dominated by the FeL complex, and the FeL peak varies over a range from roughly 0.9 keV to 1.1 keV
 as shown (in black) in figure 2. As with  all X-ray CCDs, the  EPIC/MOS instruments 
cannot resolve the FeL blends,   but the resolution is good enough to trace any  variations of  ionisation state by tracking the energy centroid of the Fe complex. Since in the temperature region of interest, 0.5-2.5 keV, the centroid of the FeL feature is a simple  function of temperature, we can easily measure small effective changes in temperature.  
Using a simple analysis of the mekal plasma model, a change in median energy 
from 0.9 to 1.1 keV for the FeL blend  corresponds  to a temperature range from 0.8-2 keV. 

To enhance the detection of the features of interest we have made a mean energy 
map (by averaging the energy of all photons falling in the same pixel) in this 
narrow (0.9-1.1 keV) energy band.
The MOS1 energy map smoothed with a Gaussian of $\sigma = 12''$ and where only pixels with at least 5 counts are taken into account, is shown in figure 3.

\begin{figure}
\begin{center}
\includegraphics[width=4.cm,height=8cm,angle=-90]{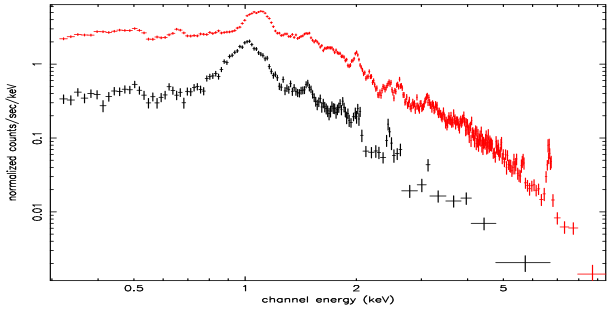}
\caption{MOS1(in black) east arm  spectrum compared with the surrounding ambient 
spectrum (in red)}
\label{figure2}
\end{center}
\end{figure}

\begin{figure}
\begin{center}
\resizebox{6.5cm}{!}{\includegraphics{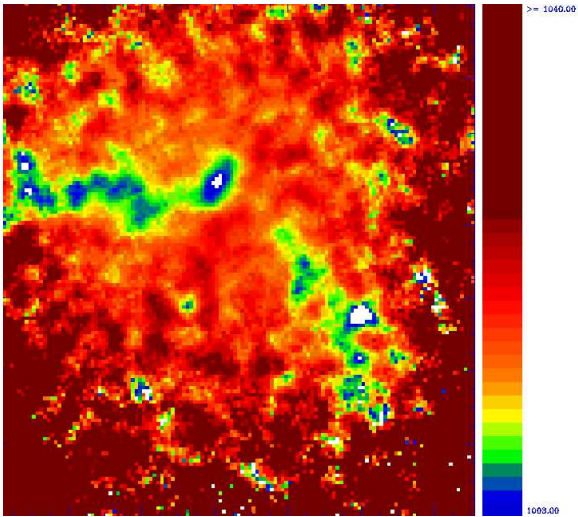}}
\caption{MOS1 0.9 -1.1 keV energy map (average of the energy of all photons 
falling in the same pixel). The image, smoothed by a Gaussian of $\sigma= 12''$, is zoomed on the regions of interest (11 arcmin field of view). The color scale is in eV.}
\label{figure3}
\end{center}
\vspace{-0.6cm}
\end{figure}

The ``arc-like structure'' revealed by the wavelets analysis is evident in more detail in this map, indicating that the  spatial structure is accompanied by spectral differences. From the X-ray center, an easterly arm which corresponds to the eastern radio lobe, extends up to 3 arcmin; the south-westerly elongation has a size of nearly 4 arcmin and appears separated from the core. Hereafter we refer to these regions as E-arm and SW-arm.  

The two arms structures show a lower mean energy than the surrounding ambient medium,  and  both the east and south-west arm present some kind of ``patchy'' structures of marginally lower mean energy than other places in the  filament: the white bubbles in figure 2 have a mean 
energy  of 1003 ($\pm5$) eV and the yellow region is at 1010 ($\pm4$) eV; the difference is significant at 1$\sigma$ statistical  errors. If this represents a true change in temperature it corresponds to a narrow variation from 1.25  to 1.35 keV.  The north-west zone in figure 3 also exhibits  apparent granularity, however the poorer statistics in this region than in the arms, suggests that this is due primarily to photon statistics noise.

\subsection {\rm Spectral analysis}

The spectra presented below are corrected for vignetting effects following the 
method presented in Arnaud et al. (2001), however  the corrections are small for the region of interest here (within 5 arcmin from the center and below 4 keV). Spectra are binned in order to achieve a  3$\sigma$ S/N ratio in each 
bin after background subtraction, and fitted using the vmekal model implemented in  XSPEC version 11. 
The response file used (on-axis matrix v13.5) is spatially uniform in the central CCD (the region of our analysis). Since the matrix is, at present, uncertain below 0.3 keV, only events above this energy are taken into account in the spectral fits.

To extract the spectrum of each arm, and to rise the S/N ratio with respect to the analysis presented in section 3.1, we have defined a region 
outlined by the 1.011 keV contour on our energy map, where the gradient is large. Since PSFs are a slightly  different for the two MOS cameras, the regions from which counts are extracted  differ slightly,  resulting in different spectral normalizations. 

We have analyzed the spectra in the arm regions using three  different methods of background subtraction:
 (i) a circular region ($R\sim$ 4.5 arcmin) centered in the middle of  
each arm (1.8 arcmin E and  2 arcmin SW from the center for the E and SW arm respectively), the arm itself being masked with a rectangle of twice its size; (ii) the signal from rectangles of 2$\times$4 armin size at a distance of 2 arcmin  from the arms, located respectively in the north and south direction (E-arm) and in the east and north-west direction (SW-arm); (iii) a circle of 4.5 arcmin in radius centered on the peak of the X-ray emission, excising the arms  and the nuclear 
region. The results are similar in all cases and we use hereafter the method (i) which defines also the 
extraction regions for the ambient medium spectra.

For the ambient medium surrounding the X-ray arms, background estimates are 
obtained from a background event list generated by summing together several 
observations in the XMM Calibration phase after sources have been excised from 
the field of view  (D. Lumb, private communication).

The background-subtracted spectra of the  eastern  and south-western arms were 
fitted with vmekal model. The free parameters were N${\rm _H}$, O, Ne, Mg, Si, 
S, Fe abundances and temperature.
A single temperature  model gives a reduced $\chi^2$= 1.8 (E-arm) and  $\chi^2$= 1.3 (SW-arm), with a 
best fit  temperature of k$T$= 1.2 keV for both arms. A similar analysis has been carried out with pn data with a very similar fitted temperature of  k$T=$ 1.35 keV, $\chi^2$/d.o.f.= 268/198 for the
E-arm  and k$T=$ 1.27 keV, $\chi^2$/d.o.f.= 245/163 for the SW one. 
 In figure 4 we show the MOS E-arm spectrum and folded 1$T$ model. Since significant residuals remain at FeL complex position, we have tried various combinations of models to test the robustness of the low temperature determination.
For both the E-arm and SW-arm regions, adding a second temperature component significantly improves the $\chi^2$ value of the fit. The statistical results of this analysis are summarized in table 1.

\begin{figure}
\begin{center}
\includegraphics[width=4.cm,height=8cm,angle=-90]{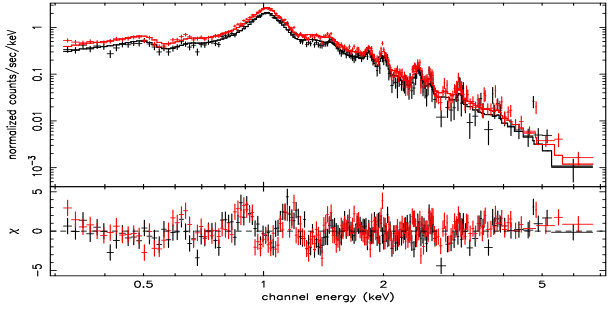}
\caption{The MOS E-arm spectrum and folded 1$T$ model. Significant residuals are present at the energy of the FeL blend}
\label{fig4}
\end{center}
\vspace{-0.6cm}
\end{figure}

\begin{table}[h]
\caption{Fitted temperature and $\chi^2$ value for the 1$T$ and 2$T$ model. The parameter of the F-test is listed in column (4) the $\Delta \chi^2$ in column (5).}
\begin{center}
\begin{tabular}{|c|c|c|c|c|}
\hline
region  & T         & $\chi^2$/d.o.f. & F-test & $\Delta \chi^2$\\
 model & (highest) &                 &  value &  \\
\hline
E-arm (1T) & 1.28 & 687/373 & &\\
E-arm (2T) & 1.64 & 592/372 & 59.7 & 0.25\\
\hline
SW-arm (1T) & 1.23 & 405/315 &  &\\ 
SW-arm (2T) & 1.51 & 318/314 & 86 & 0.27 \\
\hline
\end{tabular}
\end{center}
\vspace{-0.5cm}
\end{table}
The low-temperature component is approximately  0.9 keV whereas the second  k$T$ fitted value is higher than in the case of 1 temperature model, reaching 1.5-1.6 keV.
Although the very high signal to noise and the F-test (table 1) support the addition of the second thermal
 component, we cannot exclude that part of the poor fit with 1$T$ model was due to incompleteness in the Fe line model in the atomic physics code  around the FeL complex.

To investigate if IC emission is present in the arms regions, and if it is linked to the power emitted by the radio jet (Harris et al. 1999), we fitted the spectrum with a power law whose spectral index $\alpha$ was in the range from  1. to 2.5  (B\"ohringer et al. 2001).
The flux of the allowed  power-law component is  less than 4$\times 10^{-14}$ ergs cm$^{-2}$ s$^{-1}$  in the 0.5-8 kev band, which represents less than  1$\%$ of the flux from the thermal components. 

A 2-temperature model does not give a significantly better fit in the case of the surrounding medium spectrum (F-test value $\sim$ 5). We can explain this result considering that the extracted 
region lies roughly between 1.5 and 5.5 arcmin from the core, over which the effective temperature in M87 ( B\"ohringer et al. 2001)) changes considerably, so we expect a  range of temperature rather than 2 distinct values.

\noindent The best-fit parameters for the 2$T$ model (arms) and 1$T$ model (ambient medium) for the MOS spectra are listed in table 2. Figures 5 and 6 show the results. Similarly Figure 7 shows the results for the ambient medium surrounding the X-ray arms.

\begin{figure}
\begin{center}
\includegraphics[width=4.cm,height=8cm,angle=-90]{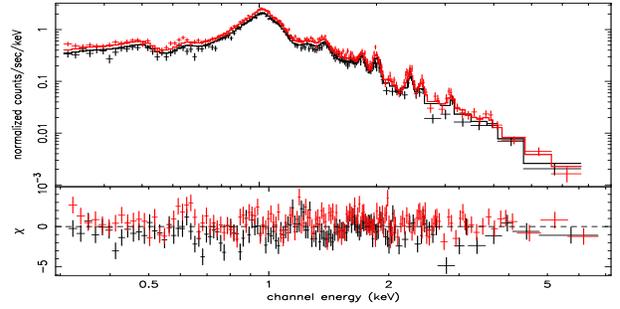}
\caption{MOS1(black) and MOS2(red) east arm  spectrum and folded 2$T$ model}
\label{figure5}
\end{center}
\end{figure}

\begin{figure}
\begin{center}
\includegraphics[width=4.cm,height=8cm,angle=-90]{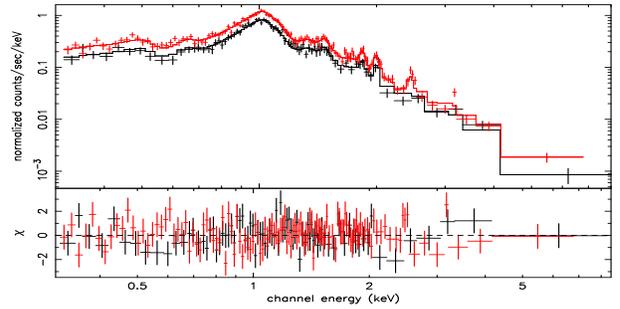}
\caption{M1(black) \& M2(red) south-west arm spectrum and folded 2$T$ model}
\label{figure6}
\end{center}
\end{figure}

\begin{figure}
\begin{center}
\includegraphics[width=4.cm,height=8cm,angle=-90]{XMM48_f7.epsi}
\caption{M1(black) and M2(red) ambient medium spectrum and folded 1$T$ model}
\label{figure7}
\end{center}
\vspace{-1.cm}
\end{figure}

\begin{table*}
\caption{Parameters of spectral fits. The first column gives the region in which the spectrum has been extracted. In columns (2) and (3) are the best fitted temperatures in units of keV (2$T$ model used in the arms regions). In column (4) is the H column density fitted value, in units of 10$^{20}$ cm$^{-2}$. Columns (5) to (7) give the fitted abundances of O, Si, S, Fe respectively. In column (8) the X-luminosity in the range 0.5-7 keV (in ergs s$^{-1}$) and the reduced $\chi^2$ in column (9). The errors correspond to 90$\%$ confidence level.}
\begin{center}
\begin{tabular}{|c|c|c|c|c|c|c|c|c|c|}
\hline
region & kT1 & kT2  &    N$_H$              & O  & Si & S & Fe & L$_x$    & 
$\chi^2$/d.o.f.\\
\hline
E-arm & 1.64 $_{-0.04} ^{+0.08}$ & 0.90 $_{-0.01} ^{+0.1}$  & 3.2 $_{-0.6} ^{+0.3}$& 0.25 $_{-0.1} ^{+0.1}$ & 1.37 $_{-0.2} ^{+0.2}$ & 1.19 $_{-0.16} ^{+0.20}$ & 0.90 $_{-0.1} ^{+0.1}$ & 0.45 10$^{42}$ & 
592/372\\
\hline
E-ambient & 2.26 $_{-0.02} ^{+0.02}$  & --- & 2.3 $_{-0.15} ^{+0.15}$ & 0.22 $_{0.05} ^{0.04}$  & 0.83 $_{-0.05} ^{+0.05}$  & 0.62 $_{-0.05} ^{+0.05}$ & 0.53 $_{-0.02} ^{+0.02}$  & 0.23 10$^{43}$ 
&904/799\\
\hline
SW-arm & 1.51 $_{-0.1} ^{+0.1}$ & 0.88 $_{-0.03} ^{+0.02}$ & 4.0 $_{-1.2} ^{+0.7}$  & 0.26 $_{-0.12} ^{+0.15}$  &  1.05 $_{-0.11} ^{+0.27}$  & 0.73 $_{-0.2} ^{+0.2}$  & 0.62 $_{-0.1} ^{+0.15}$  & 0.23 10$^{42}$ & 
318/314\\
\hline
SW-ambient & 2.28 $_{-0.03} ^{+0.02}$ & --- & 2.3 $_{-0.16} ^{+2.16}$  & 0.22 $_{-0.04} ^{+0.05}$  &  0.99 $_{-0.05} ^{+0.06}$ & 0.68 $_{-0.06} ^{+0.06}$  &   0.57 $_{-0.02} ^{+0.02}$ & 0.21 10$^{43}$ & 
1134 /805 \\
\hline
\end{tabular}

\end{center}
\end{table*}

\section{Discussion}

\subsection{Radio X-ray interactions}

The relationship between radio and X-ray is a key to understanding the  physics 
responsible for  the X-ray features (see Churazov et al. 2000; Owen et al. 2000; Harris et al. 2000). Using the best astrometry available within the SAS software, we are 
able to superimpose, in fig. 8, the contours of our energy map onto 
the 90 cm radio map, kindly provided by F. Owen (Owen et al. 2000).  
The XMM/Newton astrometry could be slightly inaccurate (estimated error on the roll angle  $<2\deg$; Watson et al. 2001, Hasinger et al. 2001) 
at this stage of the calibration, but to superimpose both arms onto their respective radio lobes it would be necessary to rotate the XMM image by 10 degrees, which is far too large to be acceptable. 

Except for the region in the E-arm described as the eastern ``ear'' (Harris et al. 2000) where
 we find good
 coincidence between our contours and Owen's map, the overlay clearly shows that the ionisation
 structure of the  X-ray arms (as traced by the mean energy map) 
 does not exactly lie on the radio lobes.
In conclusion, we can affirm that the low-temperature region  of the  X-ray arms is better correlated with the maximum brightness gradient of the lobes than  their maximum brightness.

\begin{figure}
\begin{center}
\resizebox{7.cm}{!}{\includegraphics{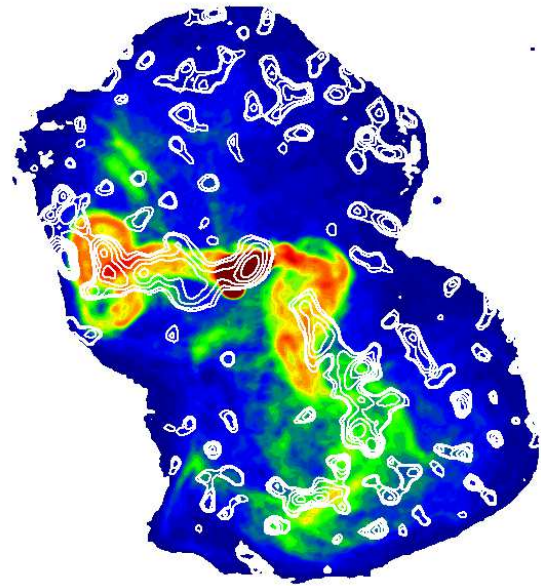}}
\caption{The radio map at 90cm, kindly provided by F.Owen, is shown with superimposed contours of MOS1 energy map. The contours are at [1000,1002,1004,1006,1008,1010,1011] mean energy level and the image as been smoothed with a Gaussian of $\sigma=40''$.}
\label{figure8}
\end{center}
\vspace{-0.6cm}
\end{figure}

\subsection{Temperature structure}

Our most significant result is the difference between  the temperature  of the 
arms and that of the  surrounding medium. Adopting the 2$T$ model the highest 
temperature of the  X-ray arms is 70$\%$  lower than the surrounding 
medium.  
The temperature of the surroundings corresponding to the E-arm and SW-arm are in 
very good agreement between each other and with the overall radial profile of M87. 

We have a strong upper limit on any non-thermal component to the X-ray  spectra of 
the lobes and thus we rule out the suggestion by Feigelson et al. (1987) that  the 
inverse Compton of the cosmic microwave background is the dominant mechanism for 
the X-ray emission from the filaments. Using equation (2) of Feigelson et al. 
we can set a lower limit to the magnetic field
of ~0.5-2. $\mu$G. This
lower limit is comparable in magnitude to Feigelson et al's
rather uncertain estimates of the equipartition field strengths
in different parts of the radio structure. A more detailed comparison
of the X-ray and radio maps is required to improve our limits on
the magnetic field strength at different positions within the lobes.

The absence of Fe K$_\alpha$ emission from the arms is noticeable in figure 2
since, at the same distance from the center,  Fe K$_\alpha$  is clearly detected in the ambient medium. However, since the statistics are poorer in the arms  than in the surroundings, we can not determine if this result comes from the lower emissivity of the  Fe K$_\alpha$ line at a temperature of 1.5 keV or it is due to the sensitivity of the instrument.

The lower temperature of the X-ray arms with respect to the surrounding medium can be explained by the mixing of plasma 
from the radio source and surroundings gas (B\"ohringer et al. 1995) and in the framework of the Churazov et al. (2000) model: the hot `radio emitting' plasma 
bubble, in nearly pressure equilibrium with the thermal gas, expands 
adiabatically and rises in the gravitational potential. During the expansion and the torus transformation (Kelvin-Helmholtz instability), it drags some colder material from the inner regions and it convects until its density equals the 
local mean density. The gas velocity is  sub-sonic and the full 
process occurs on  a time scale of a few  $\sim$ 10$^7$ years. Churazov et al. do not  consider radiative cooling in their calculations because of its longer time-scale in this region. But given the high density of the gas, radiative cooling almost surely occurs. This will explain 
the improvement to the $\chi^2$ by  the second temperature component. 
Bubbles at different radii throughout the arms are in different hydro-dynamical state. In this framework we speculate  that the gas in the arms has a complex multi-phase structure which produces the  complex temperature structure   seen  in our energy map.

\subsection{Abundances}

The abundances found in the two surrounding regions used for  background 
subtraction are in agreement with  those
found by B\"ohringer et al. (2001) at the corresponding distance from the 
center, this is not surprising even if the used extraction method is slightly 
different.

Although the fitted temperatures are the same for the two arms, within
statistical errors, the 
abundance parameters present a more complex behavior.
In the case of the E-arm, the abundance values of Si, S and Fe are significantly higher with respect to the surroundings. This is marginally true for the SW-arm. However since the abundance values depend sensitively on the exact thermal model used, this should be taken with some caution.
The higher  abundances measurement in the E-arm can be explained if, as 
proposed by Churazov et al. (2000), the uplifted gas originated in the inner 
regions where abundances are higher (B\"ohringer et al. 2001
) . However in this 
picture the lower abundances in the SW-arm are hard to understand. In an 
alternative scenario proposed by  Harris et al. (1999) the SW arm could be due 
to the shock related to the merger of the M86 subgroup with  the M87 group.

\section{Conclusion}

XMM/Newton observations of the X-ray extended features of M87 have definitively 
established the thermal nature of the plasma, with the FeL complex dominating 
the emission. The temperature in these regions is significantly lower than the 
surrounding medium. This result is very robustly established thanks to the high 
sensitivity of the experiment. The model of  
Churazov et al.  seems to qualitatively explain our result for the 
eastern arm but fails to describe the misalignment between X-ray and radio 
lobes.

\begin{acknowledgements}
We thank the XMM software team for providing the Software Analysis System (SAS) 
for the XMM-Newton data reduction. EB and JLS thank R. 
Gastaud for providing additional software for MOS data analysis. We are grateful 
to M. Arnaud for scientific discussion.
The paper is based on observations obtained with XMM-Newton, an ESA science 
mission with instruments and contributions directly funded by ESA Member States 
and the USA (NASA)"
EPIC was developed by the EPIC Consortium led by the Principal
Investigator, Dr. M. J. L. Turner. The consortium comprises the
following Institutes: University of Leicester, University of
Birmingham, (UK); CEA/Saclay, IAS Orsay, CESR Toulouse, (France);
IAAP Tuebingen, MPE Garching,(Germany); IFC Milan, ITESRE Bologna,
IAUP Palermo, Italy. EPIC is funded by: PPARC, CEA,CNES, DLR and ASI.
\end{acknowledgements}

\end{document}